# Tip Pressure Induced Incoherent Energy Gap in CaFe$_2$As$_2$

J. -X. Yin[1,2], J. H. Wang[2], Z. Wu[2], A. Li[2,4], X. J. Liang[1], H. Q. Mao[1], G. F. Chen[1], B. Lv[2], C. -W. Chu[2], H. Ding[1,3] and S. H. Pan[1,2,3]*

[1]Institute of Physics, Chinese Academy of Sciences, Beijing 100190, China

[2]Department of Physics & Texas Center for Superconductivity, University of Houston, Houston, Texas 77004, USA

[3]Collaborative Innovation Center of Quantum Matter, Beijing, China.

[4]Shanghai Institute of Microsystem and Information Technology, Chinese Academy of Sciences, Shanghai 200050, China

*Corresponding author: span@iphy.ac.cn

**In CaFe$_2$As$_2$, superconductivity can be achieved by applying a modest c-axis pressure of several kbar[1][2]. Here we use scanning tunneling microscopy/spectroscopy (STM/S) to explore the STM tip pressure effect on single crystals of CaFe$_2$As$_2$. When performing STM/S measurements, the tip-sample interaction can be controlled to act repulsive with reduction of the junction resistance, thus to apply a tip pressure on the sample. We find that an incoherent energy gap emerges at the Fermi level in the differential conductance spectrum when the tip pressure is increased. This energy gap is of the similar order of magnitude as the superconducting gap in the chemical doped compound Ca$_{0.4}$Na$_{0.6}$Fe$_2$As$_2$ and disappears at the temperature well below that of the bulk magnetic ordering. Moreover, we also observe the rhombic distortion of the As lattice, which agrees with the orthorhombic distortion of the underlying Fe lattice. These findings suggest that the STM tip pressure can induce the local Cooper pairing in the orthorhombic phase of CaFe$_2$As$_2$.**

Superconductivity in RFe$_2$As$_2$ (R = Ca, Sr, and Ba) emerges with chemical doping or physical pressure[3]. BaFe$_2$As$_2$ and SrFe$_2$As$_2$ can reach the highest T$_C$ as 38K by chemical doping and their critical pressure for superconductivity is as high as 40-60kbar[4]. In contrast, CaFe$_2$As$_2$ can be tuned into a superconductor with T$_C$ up to 49K through rare earth element doping[5]. Interestingly, applying hydrostatic pressure cannot tune CaFe$_2$As$_2$ into a superconductor[6], while superconducting critical temperature exceeding 10K can be achieved by applying a c-axis pressure as low as 5kbar[1,2]. Moreover, in addition to the tetragonal phase and the orthorhombic phase, with the smallest unit cell volume in 122 families, CaFe$_2$As$_2$ can also be tuned into the collapsed tetragonal phase by applying pressure[7]. The pressure-induced superconductivity and the collapsed phase, together with the other two phases usually coexist at the mesoscopic level[8-11]. Accordingly, there are intense debates on which phase the pressure-induced superconductivity emerges.

Tantalized by the aforementioned novel phenomena induced with modest pressure in CaFe$_2$As$_2$, we decide to study the STM tip pressure effect on this material, as the atomically sharp STM tip can introduce a significant pressure. In our experiments, single crystal samples are cleaved in cryogenic ultra-high vaccum and imaged with tungsten tips. Single crystalline samples of BaFe$_2$As$_2$, SrFe$_2$As$_2$, and Ca(Na)Fe$_2$As$_2$ used in our experiments are grown by using the self-flux method. The cleaved crystal of CaFe$_2$As$_2$ often exposes the As surfaces[12][13]. Figure.1(a) shows a group of typical line-cut differential conductance spectra taken on As-dimer-row surface of CaFe$_2$As$_2$ at 4.2K, which reveal an incoherent gap opened at the Fermi level. Since the bulk sample is deeply in the antiferromagnetic spin density wave (SDW) state at 4.2K, we first discuss whether the gap could be related to the SDW. Theoretically, an SDW gap generally does not require the particle-hole symmetry with respect to zero-energy[14], and it would be unusual for an

SDW gap to open just at the Fermi level. Moreover, from the temperature dependence of the spectra shown in Fig. 1(b), we find that the incoherent gap feature disappears at 18K, well below the bulk SDW temperature ($T_{SDW}$=150K). Therefore, the particle-hole symmetry nature of the gap structure and its much lower disappearing temperature than $T_{SDW}$ basically exclude its origin from the SDW.

In fact, such gap structure is quite unusual among parent 122 systems. In our scanning tunnelling microscopic/spectroscopic (STM/S) studies of $BaFe_2As_2$ and $SrFe_2As_2$, as displayed in Fig. 1(c), no such spectral feature is captured. On the other hand, when compared it with the coherent superconducting gap structure of $Ca_{0.4}Na_{0.6}Fe_2As_2$ ($T_C$=33K)[15] as shown in Fig. 1(c), we find that their gap value is of the same order of magnitude. Based on these comparisons and considering that the critical pressure for the emergence of superconductivity in $CaFe_2As_2$ is much lower than those in $BaFe_2As_2$ and $SrFe_2As_2$, we speculate that the tip induced gap is of the pressure-induced Cooper pairing.

The effect of tip pressure on the local structure, particularly in the c-axis, can be visualized at a step edge that breaks the translational symmetry. Here we make the scanning tunneling across an atomically sharp step edge shown in Fig. 2(a). When we reduce the tunneling resistance (as reduce the tip-sample distance), the measured step edge height decreases as shown in Fig. 2(b). This observation qualitatively agrees with our expectation that reducing the tunneling resistance effectively increases the tip pressure.

If the tip pressure induces the Cooper pairing as we speculated, the gap size should have a noticeable dependence on the tip pressure. In Fig. 2(c) we show the spectra taken with different junction resistances. They demonstrate a strong dependence of the gap size on the junction resistance. Especially, the gap magnitude initially increases, then decreases with reducing the resistance as illustrated in the inset of Fig. 2(c). It is clear that the tip sample interaction indeed affects the local structure of the $CaFe_2As_2$ material. Such dependence of the gap size on the tip pressure further supports the pressure-induced Cooper pairing scenario.

In addition to the superconductivity, a collapse tetragonal phase can also be introduced by applying pressure in $CaFe_2As_2$[7]. There have long been debates on whether such pressure induced superconductivity arises in the collapse tetragonal phase[7], the tetragonal phase[9], or the orthorhombic phase[9-11]. To find with which phase the tip pressure induced Cooper pairing is related, we examine the symmetry of the lattice. Fig. 3(a) shows a carefully calibrated high resolution topographic image of the As surface. The fast fourier transform (FFT) of this image shows sharp brag peaks (inset of Fig. 3(a)), from which we can determine the key lattice parameters: $a_{As-As}$=3.95Å, $b_{As-As}$=3.90Å and rhombic angle $\theta$=93.4°. It is clear that the As lattice undergoes a very weak orthorhombic distortion while a strong rhombic distortion. Assuming that the under Fe lattice matches the lattice distortion of the surface As lattice as illustrated in Fig. 3(b), the Fe lattice would have undergone a strong orthorhombic distortion (($a_{Fe-Fe}-b_{Fe-Fe}$)/($a_{Fe-Fe}+b_{Fe-Fe}$)=($\tan(\theta/2)-1$)/($\tan(\theta/2)+1$)=3.0%). It is important to note that, according to our experimental observations, this orthorhombicity value does not depend on the junction resistance, and such a value also agrees with our FFT study of $BaFe_2As_2$ material and with the previous STM study on the deeply under-doped Co-Ca122 material[16]. Therefore, the tip pressure induced local pairing must have arisen from the orthorhombic phase.

Finally, we point out several issues regarding to the tip pressure-induced Cooper pairing in the Orthorhombic phase. First, as the STM tip presses the material at an atomic scale, the induced Cooper pairing has to be a very local phenomenon. In addition, the non-monotonic variation of the pairing gap magnitude with different tip pressures essentially agrees with the experimental finding that there is quite generally an optimum Fe-As bonding angle for iron based superconductivity[17], which further suggests that the local Fe-As interaction controls the Cooper pairing strength. Secondly, the lack of well-defined coherent peaks in the tunneling spectra would suggest a weak phase coherence of the Cooper pairing. As the tip pressure is very local, the number of the total paired electrons is limited, therefore the phase fluctuation would be very strong. Thirdly, as mentioned previously, though the tip pressure-induced incoherent gap is compatible with the orthorhombic structural phase, there is no clear correlation between the Cooper pairing strength and the orthorhombicity, due to the fact we observe no variation of the orthorhombicity when the tip pressure changes. It is then natural to conjecture that local pairing strength is sensitive to the Fe-As coupling determined by the Fe-As height. Lastly, we propose that the tip pressure effects can be more accurately characterized by the combination of atomic force microscopy and STM with the same junction setup.

In summary, using STM/S, we have observed an incoherent gap structure in CaFe$_2$As$_2$ in its orthorhombic phase, which is attributed to the tip pressure induced local Cooper pairing. This observation also demonstrates that the local As-Fe-As coupling determined by the Fe-As height is very crucial to the Cooper pairing strength in this material.

**Acknowledgments:**
This work is supported by NSFC (11227903), Ministry of Science and Technology of China (2015CB921300, 2012CB933000), State of Texas through TcSUH, Chinese Academy of Sciences, the Strategic Priority Research Program B (XDB07030000, XDB04040300, Y4VX092X81).

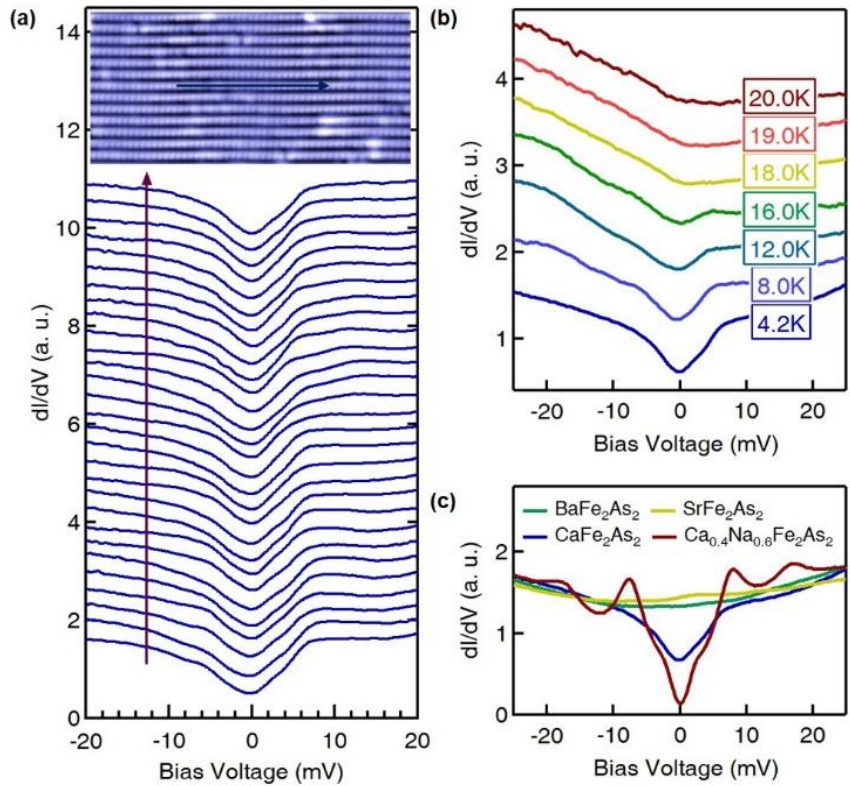

**Fig. 1** (a) Line-cut spectra taken on the As dimer-row surface of $CaFe_2As_2$ (V=-20mV, I=0.5nA, and T=4.2K). Spectra are offset for clarity. The inset displays the topographic image where the spectra are taken (V=-50mV, I=0.5nA, and T=4.2K). (b) Temperature dependence of the spectra taken on $CaFe_2As_2$ (V=-30mV, and I=1nA). Spectra are offset for clarity. (c) Comparison between spectra taken on $CaFe_2As_2$, $BaFe_2As_2$, $SrFe_2As_2$ and $Ca_{0.4}Na_{0.6}Fe_2As_2$ (V=-30mV, I=1nA, and T=4.2K).

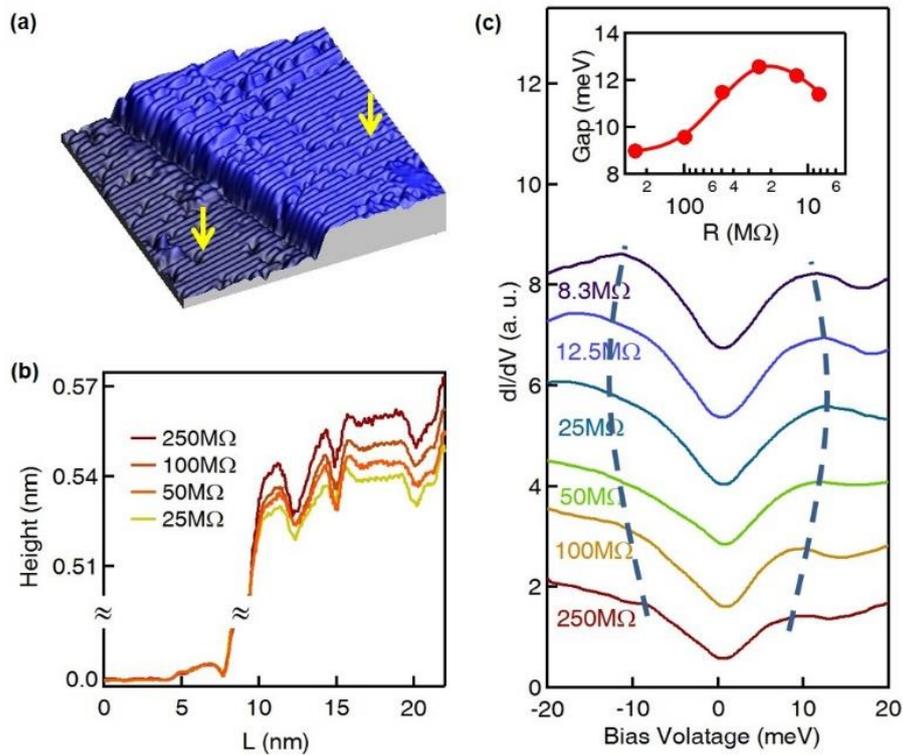

**Fig. 2** (a) Topographic image of a step edge in 3D view (V=-50mV, I=0.5nA, and 30×30nm). (b) Tunnelling junction resistance dependence of the line-cut profile across the step edge (with fixed V=-50mV and varying I). The line cut is made between the two points marked by arrows in (a). (c) Junction resistance dependence of the spectra (with fixed V=-50mV and varying I, and T=4.2K). The blue dashed lines are of guide to the eyes. Spectra are offset for clarity. The inset shows the gap size evolution as a function of the resistance. The red solid lines are of guide to the eyes. All these data are acquired at 4.2K.

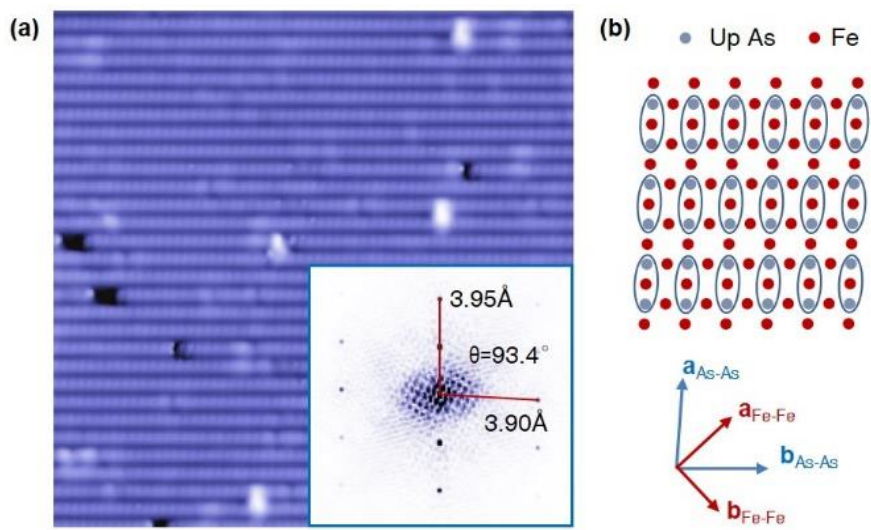

**Fig. 3** (a) Topographic image of the As surface (V=-50mV, I=0.5nA, 23×23nm, and T=4.2K). The inset shows the FFT image of the topography. By measuring the position of the brag spots we can determine the surface lattice parameters: $2a_{As-As}=2×3.95Å$, $b_{As-As}=3.90Å$, and $\theta=93.4°$. (b) Schematic diagram of the surface distortion.